\documentclass[conference]{IEEEtran}
\ifCLASSINFOpdf
\else
\fi
\usepackage{xurl}
\usepackage{flushend}
\usepackage{hyperref}
\usepackage{xcolor}
\usepackage{listings}
\definecolor{codegreen}{rgb}{0,0.6,0}
\definecolor{codegray}{rgb}{0.5,0.5,0.5}
\definecolor{codepurple}{rgb}{0.58,0,0.82}
\lstdefinestyle{mystyle}{
    backgroundcolor=\color{white},   
    commentstyle=\color{codegreen},
    keywordstyle=\color{magenta},
    numberstyle=\tiny\color{codegray},
    stringstyle=\color{codepurple},
    basicstyle=\ttfamily\footnotesize,
    breakatwhitespace=false,         
    breaklines=true,                 
    captionpos=b,                    
    keepspaces=true,                 
    numbers=left,                    
    numbersep=5pt,                  
    showspaces=false,                
    showstringspaces=false,
    showtabs=false,                  
    tabsize=2
}
\usepackage{svg}
\usepackage{comment}
\usepackage{multirow}
\usepackage{enumitem}
\usepackage{caption}
\usepackage{subfig}
\usepackage{titlesec}
\usepackage[htt]{hyphenat}
\usepackage{seqsplit}
\usepackage{soul}
\setulcolor{red}

\usepackage{xspace}
\newcommand{\name}{\textsf{\small Blindfold}\xspace}
\newcommand{\design}{\textsf{\small Blindfold}\xspace}

\newcommand{\adp}{\textsf{\small Binary Adapter}\xspace}
\newcommand{\tcb}{\textsf{\small Guardian}\xspace}
\newcommand{\secure}{sensitive\xspace}
\newcommand{\Secure}{Sensitive\xspace}
\newcommand{\shadow}{cloak\xspace}
\newcommand{\Shadow}{Cloak\xspace}

\newcommand{\extbl}{interrupt table\xspace}

\newcommand{\exvec}{interrupt vector table\xspace}

\newcommand{\shim}{\code{trampoline} segment\xspace}
\newcommand{\capy}{capability\xspace}
\newcommand{\Capy}{Capability\xspace}

\newcommand{\ptbar}{page table base address register\xspace}
\newcommand{\vbar}{interrupt table base address register\xspace}

\newcommand{\trapped}{trapped return\xspace}

\newcommand{\ie}[0]{i.e.}
\newcommand{\eg}[0]{e.g.}

\newcommand{\paragraphb}[1]{\vspace{0.05in}\noindent{\textit{#1}.}~}
\newcommand{\paragraphc}[1]{\vspace{0.05in}\noindent{\bf\em #1}~}
\newcommand{\code}[1]{\texttt{\small#1}}
\newcommand{\program}[1]{\textsf{#1}}

\usepackage{tikz}

\newcommand{\papertitle}{Blindfold: Confidential Memory Management by Untrusted Operating System}
\begin{document}
%
\title{\papertitle}

\author{
    \IEEEauthorblockN{Caihua Li}
    \IEEEauthorblockA{Yale University\\
    caihua.li@yale.edu}
\and
    \IEEEauthorblockN{Seung-seob Lee}
    \IEEEauthorblockA{Yale University\\
    seung-seob.lee@yale.edu}
\and
    \IEEEauthorblockN{Lin Zhong}
    \IEEEauthorblockA{Yale University\\
    lin.zhong@yale.edu}
}

\pagenumbering{arabic}


%


\IEEEoverridecommandlockouts
\makeatletter\def\@IEEEpubidpullup{6.5\baselineskip}\makeatother
\IEEEpubid{\parbox{\columnwidth}{
		Network and Distributed System Security (NDSS) Symposium 2025\\
		24-28 February 2025, San Diego, CA, USA\\
		ISBN 979-8-9894372-8-3\\
		https://dx.doi.org/10.14722/ndss.2025.240294\\
		www.ndss-symposium.org
}
\hspace{\columnsep}\makebox[\columnwidth]{}}

\maketitle

\begin{abstract}
    Confidential Computing (CC) has received increasing attention in recent years as a mechanism to protect user data from untrusted operating systems (OSes).
Existing CC solutions hide confidential memory from the OS and/or encrypt it to achieve confidentiality.
In doing so, they render OS memory optimization unusable or complicate the trusted computing base (TCB) required for optimization.

This paper presents our results toward overcoming these limitations, synthesized in a CC design named Blindfold.
Like many other CC solutions, Blindfold relies on a small trusted software component running at a higher privilege level than the kernel, called Guardian.
It features three techniques that can enhance existing CC solutions.
First, instead of nesting page tables, Blindfold’s Guardian mediates how the OS accesses memory and handles exceptions by switching page and interrupt tables.
Second, Blindfold employs a lightweight capability system to regulate the OS’s semantic access to user memory, unifying case-by-case approaches in previous work.
Finally, Blindfold provides carefully designed secure ABI for confidential memory management without encryption.

We report an implementation of Blindfold that works on ARMv8-A/Linux.
Using Blindfold's prototype, we are able to evaluate the cost of enabling confidential memory management by the untrusted Linux kernel.
We show Blindfold has a smaller runtime TCB than related systems and enjoys competitive performance.
More importantly, we show that the Linux kernel, including all of its memory optimizations except memory compression, can function properly for confidential memory.
This requires only about 400 lines of kernel modifications.
\end{abstract}


%

\section{Introduction}
\label{sec:intro}

\begin{figure*}[t]
    \centering
    \begin{minipage}{0.485\textwidth}
        \centering
        \includegraphics[height=0.4\textwidth]{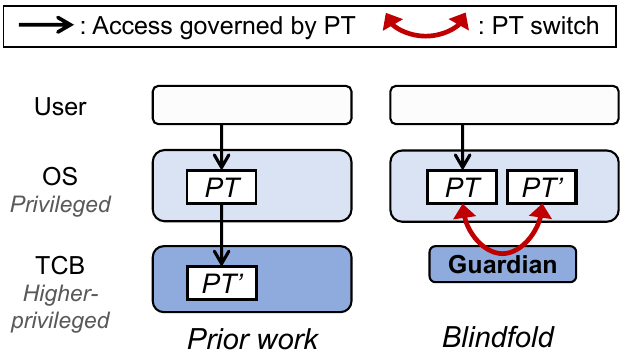}
        \caption{A very high-level comparison of \design's approach of switching vs. that of nesting used in~\cite{chen2008overshadow, hofmann2013inktag, kwon2016sego, guan2017trustshadow, li2019hypsec, li2021twinvisor, van2022blackbox}.
        In the latter, the TCB (\textit{higher-privileged}) must manage the additional level of address translations and its page tables.
        In contrast, in \design, all the page tables (PTs) are still in the OS (\textit{privileged}) while the TCB (\textit{\tcb}) only mediates their use and updates.
        Similarly \design also places the interrupt tables in the OS while the TCB only decides which one to use depending on whether the running process is \secure or not, unlike prior work in which the interrupt table is inside the TCB, \eg, ~\cite{guan2017trustshadow, li2019hypsec, li2021twinvisor}.
        See \S\ref{sec:related} for the detailed comparison.}
        \label{fig:design_comparison}
    \end{minipage}
    \hfill
    \begin{minipage}{0.485\textwidth}
        \centering
        \includegraphics[height=0.45\textwidth]{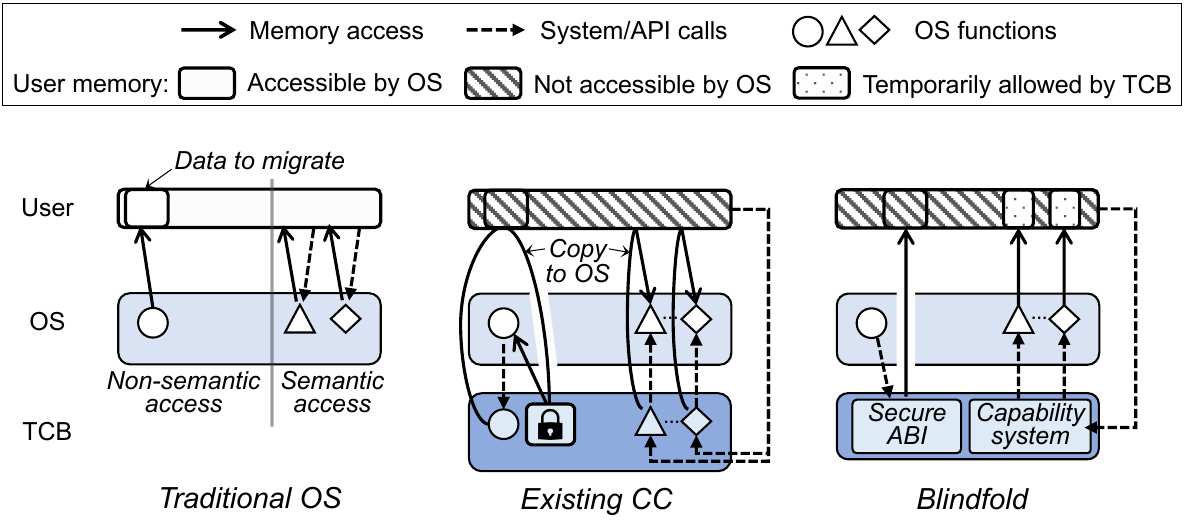}
        \caption{A comparison of \design’s techniques for \textit{semantic} and \textit{non-semantic} access vs. existing solutions.
        For non-semantic access, such as page migration, existing CC solutions either block the OS’s access to user space, disabling essential OS functions, or require expensive data copy and encryption.
        For semantic access, such as the \texttt{write} system call, existing CC solutions employ corresponding service functions inside the TCB, \ie, case-by-case, leading to an inflated TCB.
        \design resolves these issues by providing secure ABI and a capability system with general applicability.}
        \label{fig:tech_comparison}
    \end{minipage}
\end{figure*}

Modern operating systems (OSes) enjoy unfettered access to the application data.
This access is problematic because the OS may not be trustworthy, due to vulnerabilities from its large attack surface~\cite{linux_threat_2021} or lack of trust in the OS provider.
In recent years, many have attempted to ameliorate this problem under the umbrella of Confidential Computing (CC).
We say that a process or application is \secure if it does not trust the OS; we call the memory used by such applications Confidential Memory.
Existing CC solutions do not adequately support legitimate OS access to Confidential Memory and, as a result, poorly support modern big-data applications.
(\textit{i}) Some, \eg, TrustShadow~\cite{guan2017trustshadow} and BlackBox~\cite{van2022blackbox}, hide memory used by \secure applications from the OS, and as a result, OS functions stop working for such memory.
Others, \eg, Overshadow~\cite{chen2008overshadow}, resort to expensive encryption for all OS access.
(\textit{ii}) When the OS requires clear-text access to user memory, \eg, system call arguments, existing CC solutions take a case-by-case approach, leading to inflated trusted computing base (TCB) and extra data copy.
(\textit{iii}) Many of them resort to using an additional level of address translation managed by the TCB to decouple protection from address translation~\cite{chen2008overshadow, zhang2011cloudvisor, li2019hypsec, mi2020cloudvisord, li2021twinvisor, van2022blackbox}.
As a result, important OS optimization for big-data applications, \eg, page migration~\cite{linux_migration} and hugepage~\cite{linux_hugepage}, would no longer work.
(\textit{iv}) Other hardware-based solutions, such as Intel SGX, often suffer from hardware limitations. For example, applications built on top of Intel SGX suffer from a limited Enclave Page Cache (EPC) as most platforms have 128 MB or 256 MB of Processor Reserved Memory (PRM)~\cite{intel_sgx_epc_size}.

This paper reports our experience of overcoming the above limitations and allowing an untrusted Linux kernel to manage confidential memory without jeopardizing its confidentiality.
We present \design and its implementation for ARMv8-A.
Like many existing CC solutions, \design employs a small trusted software called \tcb that runs at a higher privilege level than the Linux kernel.
With \design, we demonstrate the effectiveness of a suite of techniques that can be adopted by existing CC solutions to overcome the limitations discussed above.
First, unlike previous work that deploys additional nested (or shadow) page tables and interrupt tables in TCB, \design keeps them out of the TCB (\tcb) but lets \tcb determine which one to use.
That is, it switches between these tables, instead of nesting them (see \autoref{fig:design_comparison}).
\design employs this idea to mediate memory access by the OS and the DMA (\S\ref{sec:secure_virtual_mem}) as well as protect the control flow integrity (CFI) of the protected application in interrupts (\S\ref{sec:secure_exception_handle}).
This technique can be used to isolate software regardless of which privilege mode it runs in, an objective of Tyche~\cite{castes2023creating}.
Second, unlike previous work that supports \textit{semantic} kernel access in a case-by-case manner, \design employs a single mechanism, a lightweight capability system, to support all (see \autoref{fig:tech_comparison}; \S\ref{sec:clearview}).
A \secure process explicitly grants a capability to the OS when making a system call.
Third, to enable \textit{ nonsemantic} kernel access to manage confidential memory avoiding high overhead, \design identifies the most popular memory operations in the kernel and provides secure ABI for management (see \autoref{fig:tech_comparison}; \S\ref{sec:secure_virtual_mem}), instead of always providing an encrypted view of user pages to the OS as in prior work like Overshadow~\cite{chen2008overshadow}.

With \design's prototype, we empirically show, for the first time, that the Linux kernel along with most (if not all) of its optimizations can indeed function properly with confidential memory. 
Only memory compression~\cite{linux_mem_compression} would lose its effectiveness (but still work) under \design. 
Importantly, we are able to quantify the cost of managing confidential memory.
With proper optimization, \design imposes about 3\% to 25\% performance overhead to unprotected, memory and compute intensive applications, while 10\% to 44\% to protected ones.
However, it does impose substantial overhead for I/O-intensive applications, including those that require frequent restarts.
Much of the overhead comes from cryptography and control-flow changes, instead of memory access.
Moreover, \design requires about 400 LOC of modification of the Linux kernel.
Its \tcb, implemented in mostly safe Rust (2.2K LOC), is about half the size of TCB from related systems~\cite{chen2008overshadow, guan2017trustshadow, van2022blackbox}, because \tcb does not manage memory or handle exceptions.

We note that \design only uses widely available and time-tested architectural supports (\S\ref{sec:assumption}) and as a result it is highly portable.
At the time of this writing, a basic x86-64 port (without full features) already works.
Moreover, \design supports legacy binaries and supports both protected and unprotected applications on the same system, with unprotected ones paying small performance overhead, except I/O intensive applications similar to BlackBox~\cite{van2022blackbox} and TrustShadow~\cite{guan2017trustshadow}.
\design is open-source and available from~\cite{blindfold_code} and an early prototype of it is described in~\cite{li2022mprotect}

\section{Background}
\label{sec:background}

We next provide a succinct background about OS memory management as related to \design, using Linux as a concrete example.

\paragraphb{Page Table-mediated Memory Access}
Modern systems, including the kernel and I/O devices, access memory \emph{virtually} through page table-based address translation, facilitated by the memory management unit (MMU) or IOMMU in the case of direct memory access (DMA) by I/O. 
The kernel can freely access the user space with the user page table: it enjoys the privilege of determining which page table to use and of changing a page table entry.
The kernel can also access physical frames that host \secure user data by mapping them to the kernel page table, \eg, direct mapping.
\design takes this privilege away, by trapping any kernel attempt to configure an MMU or change a page table into \tcb.

\paragraphb{Interrupt Table-mediated Control Flow Changes}
Modern systems employ an interrupt table to mediate the change in control flow.
The CPU automatically loads and executes the next instruction in memory until the current instruction triggers an exception, \eg, \code{svc} in ARM, or it receives an interrupt, \eg, I/O event.
Upon an exception or interrupt (hereafter, we simply use the term interrupt), the CPU enters the kernel mode and jumps to the corresponding entry in the interrupt table to run its handler before returning to the original flow of execution.
Because the kernel controls the interrupt table, the OS can not only access the execution context of the interrupted process but also compromise its control flow.
\tcb takes control of changing interrupt tables to protect the control flow (and execution context) in interrupts.
We note that different architectures use different names for the interrupt table, \eg, interrupt descriptor table (IDT) in x86 and exception table in ARM.

\paragraphb{Architectural Support}
Page and interrupt tables reside in memory.
When the kernel tries to access a physical frame hosting them, the MMU consults the corresponding page table entry and checks its protection bits.
By setting page table entries (PTEs) for the frames hosting page tables read-only to the kernel, \design takes away the privilege of changing the page tables from the kernel.
This is a technique widely used in the literature~\cite{dautenhahn2015nested, azab2016skee, yun2019ndss}.

The above technique must be applied with techniques that deprive kernel's privilege of arbitrarily updating virtual memory control, \eg, MMU enabled and page table base registers (PTBRs), which are \code{CR0}/\code{CR3} on x86 and \code{SCTLR}/\code{TTBR}s on ARM.
Modern architectures can trap updates of virtual memory control into a higher privilege level, \eg, by configuring \code{VMCS} on x86 and \code{TVM} in \code{HCR} on ARM, which are usually used by the hypervisor to monitor virtual machine-related events in the guest OS.
\design requires this widely available architectural support (\S\ref{sec:assumption}).

\paragraphb{OS Memory Optimizations}
Given the central importance of memory, modern OSes such as Linux feature various optimizations. 
\emph{Page migration}~\cite{linux_migration} improves memory performance of nonuniform memory access (NUMA) processors, by relocating the data closer to the processor where it will be accessed.
\emph{Demand paging} and \emph{swapping}~\cite{linux_virtual_memory} overcome the size limit of physical memory, especially in mobile and embedded systems, by moving data between physical memory and secondary storage on demand.
\emph{Huge pages}~\cite{linux_hugepage} are important for supporting big-data applications by improving the efficiency of address translation. 
Unfortunately, previous solutions~\cite{chen2008overshadow, guan2017trustshadow, brasser2019sanctuary, van2022blackbox} preclude such optimization opportunities by limiting the kernel's ability to manage the memory used by \secure applications.
It may lead to undesired consequence in real world scenarios. For example, without page swapping, the kernel can not swap out rarely used \secure pages to make space when the number of free pages is low.
As a result, it has to reject the following memory allocation requests until some \secure applications finish their computing and are willing to return the memory.

\paragraphb{Non-semantic vs. Semantic Kernel Access}
Modern OSes like Linux actively take advantage of their unfettered access to user memory.
The vast majority of cases are concerned with the kernel moving user-space data, \eg, demand paging and page migration.
Since the kernel does not need to understand the content of the data in these cases, we call such an access \emph{non-semantic}.
We note that (\textit{i}) the kernel always performs non-semantic accesses with the direct mapping in its kernel address space, instead of using the user page table, with the only exception of moving I/O data in I/O-related system calls, \eg, \code{read/write}.
(\textit{ii}) The two most popular low-level memory operations involved in non-semantic accesses are clearing a page to zero and copying a page within memory, i.e., \code{clear\_page} and \code{copy\_page} kernel functions.

In a minority of cases, such as system call arguments, the kernel does need to understand the data it accesses.
We call such access \emph{semantic}.
We note that the kernel always performs a semantic access by dereferencing a pointer in the user address space, tagged with \code{\_\_user} in the kernel source code.
Due to security concerns, such dereferences always occur in a set of narrow interfaces, \ie, \code{copy\_to\_user}, and \code{copy\_from\_user}~\cite{access_user_space, kernel_user-space_access}.
This property is one of the foundations of our solution for semantic accesses (see \S\ref{sec:insight}).

Appendix~\ref{sec:casestudy} provides a more detailed study on \emph{non-semantic} and \emph{semantic} access in Linux. 

\design deals with non-semantic and semantic accesses with different mechanisms.
It is worth noting that moving I/O data pointed to by the \code{buf} argument in \code{read/write} system calls is non-semantic by definition.
However, \design relies on end-to-end protection for I/O data (see \S\ref{sec:assumption}) and thus allows the kernel to access it in the same way as semantic access.

\section{Design Overview}
\label{sec:overview}

\begin{table*}[!t]
    \caption{Popular modern architectures support the architectural requirements of \design specified in \S\ref{sec:assumption}.}
    \label{tab:arch_req}
    \centering
    \begin{tabular}{| l | l | l | l |} 
        \hline
        \textbf{Architectural requirements} & \textbf{x86} & \textbf{ARM} & \textbf{RISC-V} \\
        \hline\hline
        Higher privilege mode than OS & \code{VMX} root mode & Hypervisor and monitor modes & Machine mode \\
        \hline
        Trapping virtual memory control & \code{VMCS} & \code{TVM} in \code{hcr} register & \code{TVM} in \code{mstatus} register  \\
        \hline
        Invoking higher privilege mode & Hypervisor call (\code{vmcall}) & Trap of cache type register (\code{ctr}) access & Environment call (\code{ecall}) \\
        \hline
    \end{tabular}
    \vspace{-1pt}
\end{table*}

In this section, we first introduce key insights behind \design design and describe its threat model and assumptions. Then we present the overview of \design in \ref{sec:design_overview}.

\subsection{Key Insights}
\label{sec:insight}

\paragraphb{Switching Instead of Nesting (\S\ref{sec:secure_virtual_mem}, \S\ref{sec:secure_exception_handle})}
To prevent the kernel from accessing sensitive memory with the user page table, \design switches page tables instead of nesting them (\autoref{fig:design_comparison}).
Since nesting requires the TCB to maintain the nested (or shadow) page tables, it suffers from two problems.
First, it increases the size, complexity, and attack surface of the TCB.
Second, by decoupling address translation (by the OS) and protection (by the TCB), the nesting invalidates the contiguity optimizations at the OS level, \eg, a huge page in the OS while small pages in higher-privilege level.
Likewise, to provide CFI in interrupts without deploying complicated interrupt tables inside the TCB, \design switches the interrupt tables, using one for all unprotected processes and another (called \textit{secure interrupt table}) for protected ones.
\design leverages a switching-based isolation method supported by modern architectures as introduced in \S\ref{sec:background}.

\paragraphb{Encrypted view and secure ABI for Non-semantic Access (\S\ref{sec:secure_virtual_mem})}
Inspired by Overshadow~\cite{chen2008overshadow}, \design provides the OS an encrypted view into the user space for non-semantic access, allowing it to keep managing confidential memory with all its optimizations, such as page swapping.
In contrast, recent systems~\cite{guan2017trustshadow, brasser2019sanctuary, van2022blackbox} chose to hide the memory regions allocated to \secure processes or containers from the OS, which can lead to unwanted consequences in real-world scenarios, as mentioned in \S\ref{sec:background}.
However, encryption/decryption is expensive despite architectural extensions for cryptography on modern processors.
As an optimization, we identify the most popular operations involved in non-semantic kernel access, i.e., clearing a page to zero and copying a page within memory, and let \tcb provide a secure ABI for such memory operations (see \S\ref{sec:secure_virtual_mem}).

\paragraphb{Capability System for Semantic Access (\S\ref{sec:clearview})}
\design employs a novel and lightweight capability system to support semantic accesses of the kernel to the user space. 
We observe that all semantic accesses share the following three properties. 
First, they are \emph{well-defined} in spatial (where) and temporal (when) boundaries.
Second, the user process knows when and where the kernel accesses its address space. 
Third, for security reasons~\cite{kernel_user-space_access}, the OS accesses the user space via a set of narrow interfaces, as in both the Linux and FreeBSD kernels~\cite{access_user_space,freebsd_copyin}.
Such interfaces are stable and can date back to the Linux kernel v2.2 and FreeBSD v2.2.1.
These properties are the foundations of our solution for semantic accesses (see \S\ref{sec:clearview}).
Unlike prior systems which handle semantic access, especially system calls in a case-by-case manner~\cite{chen2008overshadow, van2022blackbox}, \design's capability system handles all with the same design, substantially reducing the TCB size.

\paragraphb{Architecture-agnostic Design}
While recent related work has often exploited architecture-specific support, \eg, ARM TrustZone~\cite{guan2017trustshadow, brasser2019sanctuary} and Intel SGX~\cite{arnautov2016scone}, we design \design to rely on only time-tested and universally available architecture features. 
In doing so, we aim not only to widen the user base, but also to sidestep the availability and security risks often associated with new hardware features.
We report an implementation on ARMv8-A in \S\ref{sec:impl}, and discuss the portability to x86-based systems in \S\ref{sec:discussion}.

\subsection{Threat Model and Design Space}
\label{sec:assumption}

\paragraphb{Threat Model}
A \secure application trusts the hardware and the \tcb.
It also trusts the tools and libraries used by its developers.
We assume secure boot and thus the OS and the \tcb are supposed to be initialized securely.
However, we do not trust the OS or any other software at runtime, which means that the OS may be compromised after booting.

We protect the confidentiality and integrity of application data against any adversaries that can compromise the OS or access memory via DMA.
We also protect the code integrity of the application and control flow integrity (CFI) across user-kernel interface, i.e., exceptions and interrupts, to fend off attacks such as return-oriented programming.

We do not protect the data sent to or received from out of a process, such as I/O and inter-process communication. Like HypSec~\cite{li2019hypsec} and BlackBox~\cite{van2022blackbox}, we believe such data is better protected end-to-end~\cite{saltzer1984end}.
We defend against replay attack only for the Guardian-encrypted data such as swapped pages by maintaining per-page signatures (\S\ref{sec:secure_virtual_mem}).
We also defend against memory mapping-related Iago attack, which means we check the return value of system calls like \code{mmap} and \code{brk}.
However, we do not defend against denial-of-service (DoS) attack. Physical and side-channel attacks are not our targets either.

\paragraphb{Constraints}
\design has two absolute constraints. 
(\textit{i}) It must support legacy binaries.
We believe protecting application data should be transparent and orthogonal to application development and should support binaries that already exist.
(\textit{ii}) Its design must be architecture-agnostic and therefore eschew features that are available only in some specific architectures, e.g., ARM TrustZone.

\paragraphb{Tradeoffs} We make two important tradeoffs in designing \design.
(\textit{i}) We balance between changes to the OS and the size of the runtime TCB (\tcb).
While it is desirable to keep both small, when we have to choose one over the other, we choose a small \tcb.
\design is able to achieve 2$\times$ smaller runtime TCB while requiring a similar amount of OS modifications compared to the state of the art.
(\textit{ii}) Since non-sensitive applications in the same systems may not require protection, they ideally should pay little or no performance penalty.
While it is desirable to keep both small, when we have to choose one over the other, we choose a small overhead for non-sensitive applications.

\paragraphb{Architectural Requirements}
\design has three requirements for the architecture.
x86, ARM, and RISC-V all meet these requirements, as summarized in \autoref{tab:arch_req}.
\begin{enumerate}[topsep=0.1em,itemsep=-0.1ex, leftmargin=*, label=\textit{A$_{\arabic*}$.}]
    \item A programmable higher privilege level than what the OS is running in.
    \item A way to trap updates of the virtual memory control into the higher privilege level.
    \item A mechanism with which an application can invoke the higher privileged software, bypassing the OS.
\end{enumerate}
Additionally, we note that software-based approaches such as Nested Kernel~\cite{dautenhahn2015nested} and SKEE~\cite{azab2016skee} offer alternatives to implement a programmable higher privilege level without reliance on specific architectural support.
These approaches can facilitate more efficient prototypes by allowing privilege level transitions to resemble kernel function calls, avoiding the overhead associated with hardware privilege level switching.

\subsection{\design Overview}
\label{sec:design_overview}

\begin{figure}[!t]
    \centering
    \includegraphics[width=0.485\textwidth]{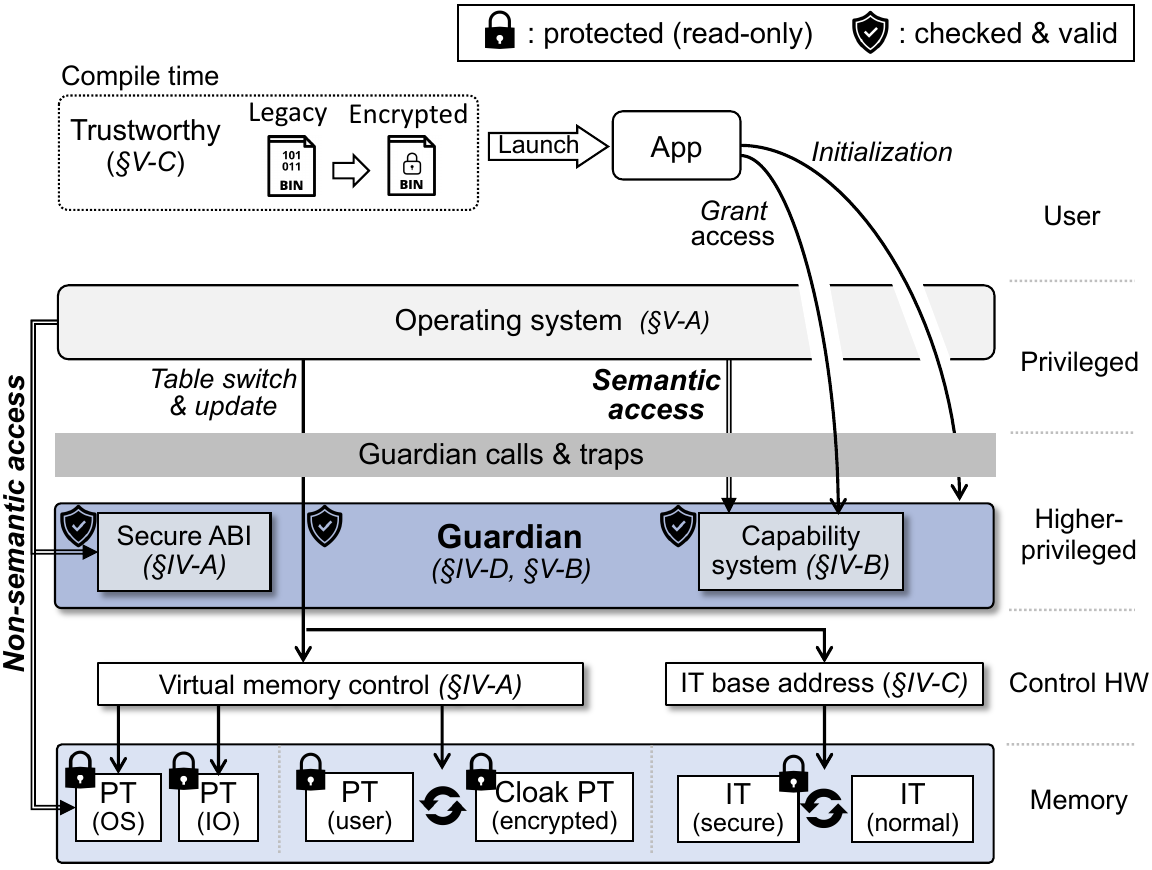}
    \caption{\design Overview.
    \design switches the page tables (PTs) and the interrupt tables (ITs) to ensure all OS memory access (\S\ref{sec:secure_virtual_mem} and \S\ref{sec:clearview}) and control flow changes (\S\ref{sec:secure_exception_handle}) are mediated by the TCB (Guardian) (\S\ref{sec:guardian_design}).
    For semantic access, we employ a lightweight \capy system in the \tcb to verify access from the OS (\S\ref{sec:clearview}).
    \design provides an interface to enable direct communication between the application and the \tcb bypassing the OS (\S\ref{sec:guardian_design}).}
    \label{fig:overview}
\end{figure}

\design protects the confidentiality and integrity of the code and data of the application and control flow integrity (CFI) across the user-kernel interface.
Its core is a small trusted software called \tcb that runs at a higher privilege level than the OS (\autoref{fig:overview}).
The \tcb implements the four key insights described in \S\ref{sec:insight}.

\tcb protects \secure user memory from kernel's unfettered accesses by enforcing the following invariants.
\begin{enumerate}[topsep=0.1em,itemsep=-0.1ex, leftmargin=*, label=\textit{I$_{\arabic*}$.}]
    \item Virtual memory is always enabled after secure boot;
    \item All page table updates must go through the \tcb;
    \item Any kernel (or colluded user) thread can never access \secure user space via a \secure user page table;
    \item Any \secure page is either encrypted, or mapped in its associated user page table exclusively (or not mapped at all as a transient state).
\end{enumerate}
With the above four invariants, \tcb guarantees that any kernel thread (or colluded user thread) can never access \secure user pages in clear text.
For semantic accesses, the OS must invoke \tcb, which verifies the capabilities before copying the clear-text data to the OS (\S\ref{sec:clearview}).

When a \secure process is running, tcb forces the system to use the secure interrupt table so that it mediates interrupt handling (\S\ref{sec:secure_exception_handle}) to ensure control flow integrity (CFI) in interrupts and protect \secure process context.
While the \tcb may resemble a hypervisor, it does not manage any resources or handle interrupts.
As a result, \tcb remains small.

In addition to \tcb, \design employs three more components to support legacy applications (see details in \S\ref{sec:impl}). 
(\textit{i}) \emph{Secure boot} during which \tcb enables the MMUs and invalidates the writable permission of the OS to all page tables;
(\textit{ii}) \emph{Binary adaptation} prepares a legacy app binary for protection, encrypting loadable segments and adding helper segments;
(\textit{iii}) A small set of OS modifications so that page table updates and legitimate semantic kernel accesses must make function calls into \tcb.

\section{Design Details}
\label{sec:details}

We next provide details for the novel design ideas of \design.
(\textit{i}) Switching page tables and providing the OS an encrypted view to regulate non-semantic kernel access (\S\ref{sec:secure_virtual_mem});
(\textit{ii}) A lightweight capability system to support OS semantic access (\S\ref{sec:clearview});
(\textit{iii}) Switching interrupt tables to protect CFI in interrupts (\S\ref{sec:secure_exception_handle}). 
Finally, we describe the \tcb interface and security analysis.

\subsection{Regulated Non-semantic Access}
\label{sec:secure_virtual_mem}
\name enables performant and secure non-semantic accesses by providing secure \tcb ABI. We first explain how \name disallows the OS from accessing user memory via its own page tables, instead creating an encrypted view of the user memory. We then describe the secure ABI design to optimize common cases.

\paragraphb{Switching between user and \shadow page tables}
In Linux, the kernel can access the user space of a process with its user page table.
\design forces the kernel to use a \textit{\shadow page table} instead, when control is transferred from the \secure process to the OS (see \S\ref{sec:secure_exception_handle}).
The \shadow page table does not map any \secure pages in the user space so the kernel cannot perform a non-semantic access via user space pointer dereference.
We note that the kernel can still function properly with the \shadow page table because it always performs a non-semantic access with the direct mapping in kernel address space (see \S\ref{sec:background}).
In other words, employing the \shadow page table will only block illegitimate non-semantic accesses via user space pointer dereference.

\paragraphb{Switching between clear and encrypted views mapped in user and kernel spaces respectively}
At any moment, each \secure page is either (\textit{i}) in clear text and exclusively mapped in the associated user space, or (\textit{ii}) encrypted, unmapped from the associated user space, and may be mapped in kernel space for non-semantic accesses (or (\textit{iii}) in clear text but not mapped in any space as a transient state for optimization introduced later in this section).

As \tcb is invoked in every update of a page table, it can decrypt (or encrypt) a \secure user page when the page is mapped to (or unmapped from) the user space.
Meanwhile, \tcb invalidates (or validates) the PTE of direct mapping in the kernel space associated with the physical frame that hosts the \secure page.
\tcb counts mappings to each physical frame in all page tables to ensure the exclusive mapping of a \secure page in clear text. The counting is enabled by the fact that all the updates to PTBR and a page table are trapped in \tcb.

When a \secure page is unmapped from user space and encrypted, the \tcb generates and stores the signature of the page in a reserved region in user space (see \S\ref{sec:impl_proc}). The signature is for verification when the page is mapped to user space again.
The kernel may swap out the pages for storing signatures, in which case, the signatures of these pages are generated, and thus form a Merkle tree lazily.
If the root of the tree is swapped out, its signature is stored in the \tcb.

We note that this design does not lead to race condition or frequent switching between two views in a non-semantic access.
When the kernel performs non-semantic access, mostly for memory management like swapping and page migration, it will first lock the page and unmap it from the user space, to prevent modification during the memory operation like page movement. As a result, the lock and the unmapping prevent any race condition between the user process and the kernel.
Such operations do not happen frequently either, assuming that the kernel performs well in its job of memory management.
However, a malicious kernel can indeed slow down the process execution by maliciously performing non-semantic accesses frequently. This denial-of-service (DoS) attack does not harm the \secure data, which we do not defend against (\S\ref{sec:assumption}).

\paragraphb{Secure ABI for memory operations}
Encryption and decryption are expensive.
Instead of always providing encrypted view of a \secure page for non-semantic accesses, we identify two most popular involved operations, \ie, \code{clear\_page} and \code{copy\_page}, and avoid unnecessary encryption/decryption when possible.

First, in cases that (\textit{i}) a \secure process actively frees a virtual memory area (VMA) and (\textit{ii}) the kernel kills a \secure process when it finishes (or for any other reason), the kernel first unmaps the associated pages from user space and then removes the pages to zero.
Without being aware that the pages are no longer used, \tcb encrypts and signs the page unnecessarily.
We optimize these cases through an ABI that allows the kernel to notify \tcb that one specific VMA or all VMAs are being freed.
We note that the kernel cannot abuse this ABI because the \tcb marks the VMAs are freed, unmaps and clears the associated pages in the ABI call. Any abuse will be detected when the \secure process tries to access the pages after the call.

Second, in case of copying/moving a \secure page from one physical frame to another, \eg, page migration, its content must be protected by encryption if the kernel performs the copying. Afterwards, the page content must be decrypted when the new physical frame is mapped to the same page.
In this case, we avoid encryption/description through an ABI that triggers \tcb to copy. \tcb generates and stores the signature of the page, checks if the target frame is already mapped in any user space, and invalidates associated kernel direct mapping before performing the copy.
This implies that the copied page is in clear text but is not mapped to any space. The kernel can later remap the page to a \secure process if and only if the target process has a matched signature.
We note that the kernel cannot abuse this ABI because only a verified process with the matched signature has the mapping and thus can access the copied page. The kernel can also not steal information from or fake the signature because the signature is not available for the kernel until it is encrypted and protected by the Merkle Tree in a swap-out.

\paragraphb{DMA access}   
To prevent illegitimate DMA access to private data of a \secure process, \design also mediates updates to the IO page table.
The \tcb only allows mappings from an IO page table to public pages, i.e., pages with \code{MAP\_SHARED} being set in a \code{mmap} system call.

\subsection{Capability-based Semantic Access}
\label{sec:clearview}
Unlike previous works in which semantic accesses are supported in a case-by-case manner, \design supports all semantic accesses with the same \capy system, without extra copying.
We design the \capy system based on the three properties shared by all semantic accesses as mentioned in \S\ref{sec:insight}.
A \capy is a conceptual representation of the permission to access a contiguous region in the user address space.
Unlike traditional \capy-based security, \design's \capy is stored inside \tcb, and no key is provided to the OS.
The OS can request read or write access with a virtual address, and \tcb checks if the corresponding \capy exists before granting access to the OS.

Each \capy is specified by a 4-tuple \code{(addr, size, rw, life)} where the elements represent the start address of the region, the size in bytes, the permission if it is read-only or read/writable for the OS, and the lifetime of the \capy, respectively.
\tcb maintains a list of capabilities for each \secure process.
In our prototype, we implement it as a sorted list based on the start address and wrap it with a readers-writer lock, so it allows for concurrent read and the search time is $O(log(n))$ where $n$ is the number of capabilities.

\paragraphb{\Capy Creation and Destruction}
\design does not require more information than the system call semantics and the arguments to create/destruct capabilities.
To be more precise, we predetermined the required capabilities based on the semantics of system calls, \eg, how to determine the range of region and read/write permissions from the system call arguments.
At runtime, the Guardian creates/destructs capabilities when a system call is trapped, based on the provided system call identifier and arguments, before/after the OS serves the system call.
Specifically, for nested data structures, \ie, pointers to pointers, \tcb reads the address and length from the user space, as indicated by the pointers in the arguments, and generates the corresponding capabilities.
Our current prototype supports up to three levels of pointers (\S\ref{sec:eval_coverage}).

\design categorizes \capy into short- and long-lived.
For almost all capabilities of which the lifetime ends with the system call returns (\ie, short-lived), the \tcb records the thread identifier (\ie, the stack pointer; see \S\ref{sec:guardian_design}) in \code{life}.
Upon return of the system call, which also traps in \tcb, \tcb destructs the \capy, \ie, removing it from the list.
We note that \code{clone}, \code{set\_tid\_address}, and \code{set\_robust\_list} are the only three system calls that can create a long-lived \capy that is alive until the end of the calling thread.
These system calls register a 4- or 8-byte region that the kernel will access exactly once when the thread terminates.
For long-lived \capy, \tcb records the base address of the stack in \code{life}.
These capabilities are disposable, which means they are destructed once they are accessed.

\paragraphb{\Capy Check and Semantic Access}
The Unix-like OSes always access user space data with care, through narrow and stable interfaces such as \code{copy\_from\_user}~\cite{access_user_space}.
\design modifies the implementation of these interfaces so that \tcb is invoked (via \code{g\_mov\_mem} in \autoref{tab:tcb_abi}) to check \capy, \ie, if any \capy matches the accessing address and the read/write permission.
Once \tcb verifies the legitimacy, it copies the data between the user and kernel space according to the request.
Since there is no extra copying as in the buffer-based approach used in prior work~\cite{chen2008overshadow, guan2017trustshadow, van2022blackbox} (\ie, user $\rightarrow$ buffer $\rightarrow$ kernel, instead of user $\rightarrow$ kernel), \design's approach is both more efficient and more general (see comparison in \S\ref{sec:related}).

\begin{table*}[t]
\caption{\tcb's ABI consists of 10 calls, which are invoked by various components in the system, including hardware, the secure \extbl, the OS, and the \texttt{trampolines} in \secure processes. These calls play crucial roles in regulating OS operations, maintaining OS functionality and supporting \secure processes.}
    \label{tab:tcb_abi}
     \centering
    \begin{tabular}{l|c|l}\hline
        \textbf{\tcb ABI}& \textbf{Where invoked}  & \textbf{Role} \\ 
        \hline\hline
        \code{g\_vmc\_trap}& Update of virtual memory control & Trap update to virtual memory control \\
        \hline
        \code{g\_interrupt}& Secure \extbl & Trap when an interrupt happens to a running sensitive process \\
        \hline
        \code{g\_set\_pt} & \multirow{5}{*}{OS} & Trap when the OS tries to update a page table \\
        \code{g\_free\_vma} &  & Trap when the OS reclaims the memory of VMAs \\
        \code{g\_copy\_page} &  & Trap when the OS copies a page within memory \\
        \code{g\_move\_umem} &  & Trap for semantic access to user space by the OS \\
        \code{g\_fork} &  & Trap when the OS forks a \secure process \\
        \hline
        \code{g\_proc\_create} & \multirow{3}{*}{\Secure process (\texttt{trampolines}) } & Trap when a \secure process is being created \\
        \code{g\_proc\_resume} & & Trap when a \secure process resumes after exception handling \\
        \code{g\_proc\_signal} & & Trap when a \secure process is ready to handle a signal \\
        \hline
    \end{tabular}
\end{table*}

\subsection{Switching Interrupt Tables for CFI in interrupts}
\label{sec:secure_exception_handle}
To protect the control flow integrity and execution context of a \secure process in interrupts and exceptions, \design makes the memory storing the interrupt tables read-only and forces the system to use a modified \emph{secure \extbl} when a \secure process is running, which invokes the \tcb before the OS gains control.
When invoked, the \tcb saves then clears the context, sets the \ptbar to the \shadow page table, sets the \vbar back to the original interrupt table, and finally forwards the control to the interrupt handler.

\design's design of switching between two interrupt tables contrasts those taken by prior work.
For example, many~\cite{chen2008overshadow,li2019hypsec,van2022blackbox} use one interrupt table in the OS to trap all interrupts into the TCB regardless of whether the running process is \secure or not.
\design's design avoids unnecessary overhead for non-\secure processes.
Ginseng~\cite{yun2019ndss} dynamically modifies the interrupt table at \emph{runtime} to invoke the TCB, incurring higher runtime overhead.

\paragraphb{Resumption with Trapped Return}
When the OS resumes the execution of a \secure process after it handles an interrupt, it simply returns control to the process starting with the saved program counter; as a result, there is no obvious point that the \tcb could intervene.
\design solves this problem with a simple mechanism called \emph{\trapped}.
When the \tcb clears the context, it sets the return address of the interrupted process to the user-space trampoline that invokes the \tcb (\code{g\_proc\_resume} in Table \ref{tab:tcb_abi}).
As a result, when the OS returns control back to the process, it unknowingly invokes the \tcb.
As such, \trapped does not require OS modification.

\subsection{\tcb Design}
\label{sec:guardian_design}
\tcb is the software TCB running in a higher privilege mode. 
It is similar to a micro hypervisor, but the \tcb does not manage any resources (including nested/shadow page tables) or handle exceptions; nor does it rely on nested paging hardware support, resulting in a smaller size.
\tcb protects itself by forbidding any mapping to the frames that host its own memory.
This is possible because in \design, the page tables are read-only to the OS and all the updates to them must go through \tcb.
\tcb has a narrow ABI as summarized in Table \ref{tab:tcb_abi}.
\tcb is reentrant and can be concurrently invoked from multiple threads running on different cores.

\paragraphb{Bookkeeping}
\tcb tracks the necessary information of the physical frames and sensitive processes, without relying on the data structures in the OS.
For each frame, it tracks whether it hosts sensitive data and a reference count across all page tables.
For each \secure process, the \tcb keeps (\textit{i}) base addresses of the user and \shadow page tables; (\textit{ii}) addresses of the user space trampolines and signature segments; (\textit{iii}) secret keys for cryptography; (\textit{iv}) \capy list for semantic access; (\textit{v}) execution context when preempted; (\textit{vi}) a list of 3-tuple \code{(start, end, status)} which represents the range and properties of virtual memory areas; (\textit{vii}) signature of the root signature page.
\tcb identifies a process by its page table base address and a thread within a \secure process by the value of the stack pointer (SP).
In particular, it does not trust the process/thread identifier (PID/TID) assigned by the OS.

\paragraphb{Memory Use}
The memory used for bookkeeping physical frames is about 16 MB in our prototype with 8 GB main memory.
That used for bookkeeping \secure processes is proportional to the number of \secure processes/threads.
This memory use is dominated by (\textit{iv}) and (\textit{v}).
Since each thread has at most two long-live capabilities, while system calls are not nested and have up to six parameters, a thread requires less than 1 KB in our prototype on ARMv8-A.

\subsection{Security Analysis and Attack Scenarios}
\label{sec:attack_surface}
We first analyze how \design achieves the security invariants listed in \S\ref{sec:design_overview} (\textit{I$_1$} to \textit{I$_4$}) and then discuss how \design protects against popular attacks.

\paragraphb{\textit{I$_1$}~Virtual memory is always enabled after secure boot}
\design assumes secure boot, during which the \tcb ensures the virtual memory is enabled and properly configures the hardware to trap updates of virtual memory control (\S\ref{sec:background} and \S\ref{sec:assumption}).
And therefore, after secure boot, the \tcb can detect any attempt of disabling virtual memory.

\paragraphb{\textit{I$_2$}~All page table updates must go through the \tcb}
According to \textit{I$_1$}, the kernel cannot bypass the virtual memory protection.
So the kernel cannot modify page tables bypassing the read-only protection as \design marks all page tables as read-only.
The kernel cannot switch page tables without going through the \tcb either, as \design traps and verifies any such attempts (\ie, updating the page table base registers; \S\ref{sec:background}).

\paragraphb{\textit{I$_3$}~Any kernel (or colluded user) thread can never access \secure user space via a \secure user page table}
The \tcb allows the use of a \secure user page table if and only if the \secure process, \ie, the owner of the data, has control.
In all other cases, every time the control changes to the kernel or another process, the \tcb switches the page table to the corresponding one other than the \secure user page table.
Whether the change of control is triggered (\textit{i}) by the \secure process explicitly, \eg, a system call, or (\textit{ii}) by an interrupt from outside the process, the \tcb always gains control before the switching via the secure \extbl and ensures that the appropriate page table is used (\S\ref{sec:secure_exception_handle}).

\paragraphb{\textit{I$_4$}~Any \secure page is either encrypted, or mapped in its associated \secure user page table exclusively (or not mapped at all as a transient state)}
All \secure pages are originally encrypted.
The \tcb ensures that only an exclusive mapping exists for a \secure page during and after transitions from an encrypted view to a clear view (and vice versa).
Specifically, the \tcb decrypts the pages only after invalidating the corresponding kernel direct mappings and ensuring that the associated frames are mapped exclusively to the \secure user space (\S\ref{sec:secure_virtual_mem}).
We note that, by leveraging \textit{I$_2$}, the \tcb maintains a complete list of any existing mappings in the page tables.
The \tcb does the opposite for the transition from a clear view to an encrypted view.
The transient state does no harm to confidentiality since it does not have any mappings or allow any access (\S\ref{sec:secure_virtual_mem}).

We note that \textit{I$_3$} and \textit{I$_4$} guarantee the kernel (and colluded user processes) can never access a \secure page in clear text.

\paragraphb{Attack Scenarios}
We discuss how \design protects against popular attacks.
We note \design does not defend against Denial-of-service (DoS) attacks as pointed out in \S\ref{sec:assumption}, \eg, slowing down the process execution by maliciously performing non-semantic access
frequently as discussed in \S\ref{sec:secure_virtual_mem}.

\textit{Replay attacks}:~~
The OS can not replay encrypted data since the \tcb maintains and checks the per-page signatures (\S\ref{sec:secure_virtual_mem}).
Even if the OS swaps out a page containing the signatures, the \tcb encrypts it and generates signature of the encrypted page, effectively forming a Merkel Tree.

\textit{Return-oriented programming (ROP) attacks}~\cite{roemer2012return}:~~
The OS can modify the return address to bypass \trapped (\S\ref{sec:secure_exception_handle}) and perform a ROP attack.
However, the compromised process immediately ``loses'' its sensitivity --- it can no longer access the sensitive context or switch to its user page table. As a result, it will not be able to access its own memory.
That is, even a successful ROP attack will never compromise any sensitive data.

\textit{Iago attacks}~\cite{checkoway2013iago}:~~
The OS may misbehave in interrupt and system call handling.
For example, it may return the address of the stack in a \code{mmap} system call, tricking the process to overwrite its stack unknowingly.
\design fends off such memory-mapping Iago attacks by maintaining its own bookkeeping of memory frames and per-process virtual memory areas.
Like BlackBox~\cite{van2022blackbox}, \design defends against a stronger threat model than TrustShadow~\cite{guan2017trustshadow}, which protects both sensitive pages and page tables by hardware.
\section{Implementation}
\label{sec:impl}

To validate \design, we prototype it with Linux kernel v5.15 on ARMv8-A architecture.
Specifically, we implement the \tcb on top of ARM Trusted Firmware~\cite{armtf} for ARMv8-A.
We note that \design has a smaller, \ie, about a half of, runtime TCB and requires a similar amount of kernel modifications compared to recent related systems~\cite{guan2017trustshadow, van2022blackbox} that also support legacy binaries.
Beyond the kernel modifications (\S\ref{sec:impl_os}) and the \tcb (\S\ref{sec:impl_tcb}), \design requires binary adaptation to support legacy binaries (\S\ref{sec:impl_proc}).

\subsection{Linux Kernel Modification}
\label{sec:impl_os}

\design introduces about 400 LOC to the Linux kernel, which is close to that in TrustShadow~\cite{guan2017trustshadow} (0.3K) and BlackBox~\cite{van2022blackbox} (0.5K).
Next, we describe some key changes.

\paragraphb{Trapping Page Table Updates}
\design requires that all page table updates be trapped into the \tcb.
To achieve this, \design modifies the implementation of low-level stable kernel interfaces like \code{set\_pte} to add trampolines to the \tcb via ABI \code{g\_set\_pt}.
We note these interfaces have been stable since Linux kernel v2.0 and the modification is about 50 LOC.

\paragraphb{\Shadow Page Table for Blocking User Space Pointer Dereference}
\design disallows the kernel to directly dereference a user space pointer in order to access \secure user pages.
To achieve this, \design modifies the kernel in three places to leverage the \shadow page table.
First, we add a new field, \code{c\_pgd}, to the memory descriptor \code{mm\_struct} to store the base address of the table of pages \shadow.
Second, we modify the kernel to create (destroy) a \shadow page table for a \secure process when the process is created (terminated).
When \tcb is invoked by \code{g\_proc\_create} and \code{g\_fork}, it ensures that the \shadow page table is read-only for the kernel, before allowing the \secure process to start (\S\ref{sec:impl_tcb}).
Finally, we modify the page fault handling logic so that it invokes the \tcb through \code{g\_set\_pt} to update the \shadow page table if the page fault happens in the \code{trampoline} segment (see \S\ref{sec:impl_proc}), which is the only page that is mapped in the \shadow page table and invokes the \tcb in \emph{\trapped} (\S\ref{sec:secure_exception_handle}).
These involve modification of about 40 LOC.

\paragraphb{Secure ABI for Optimizing Confidential Memory Management}
To reduce the overhead of unnecessary encryption/decryption for confidential memory management, we modify kernel functions such as \code{unmap\_vmas}, \code{migrate\_page\_copy} and \code{do\_cow\_fault} to invoke \tcb via \code{g\_free\_vma} and \code{g\_copy\_page}.
So, the kernel does not actually perform the non-semantic access via low level interface like \code{clear\_page} and \code{copy\_page}.
Instead, \tcb does it on behalf of the kernel.
This incurs about 60 LOC of kernel modification.

\paragraphb{Semantic Access}
To semantically access user data, we modify the implementation of intra-kernel interfaces like \code{copy\_from\_user}~\cite{access_user_space}, which have been stable since Linux kernel v2.2.
The modification invokes \tcb with \code{g\_mov\_umem} to check \capy and then perform the copy on behalf of the kernel (\S\ref{sec:clearview}).
This incurs about 40 LOC of kernel modification.

\vspace{2pt}
Beyond the above modifications, we also implemented an optimization to group small protected pages into a few huge pages, to minimize the impact of breaking down the kernel direct mapping in huge pages. The rest of the modification is mainly comprised of a general interface to make Guardian ABI call written in inline assembly and some macro defined in header files.

\subsection{\tcb Implementation}
\label{sec:impl_tcb}

We implement \tcb on top of the ARM Trusted Firmware~\cite{armtf} that runs in EL3 on ARMv8-A.
Our prototype consists of about 2.2K LOC in Rust for high-level logic and about 0.2K lines of assembly for switching between worlds.
The \tcb additionally employs the cryptography libraries (13K LOC) from the RustCrypto project~\cite{rust_crypto}, \code{linked\_list\_allocator} crate (1.2K LOC) and the readers-writer lock from the \code{synctools} crate (0.7K LOC).
We next describe a few key implementations.

\paragraphb{Page Table Control}
\design traps updates to (\textit{i}) the virtual memory control and (\textit{ii}) the page tables into the \tcb.

For (\textit{i}), the \tcb sets the \textit{TVM} bit in \code{HCR\_EL2} on ARMv8-A (\autoref{tab:arch_req}).
As a result, when the kernel updates virtual memory control registers such as \code{SCTLR\_EL1} and \code{TTBRx\_EL1}, the hardware will trap into EL2 where \tcb ABI \code{g\_vmc\_trap} (\autoref{tab:tcb_abi}) will be invoked.
\tcb updates the virtual memory control on behalf of the kernel after checking that the update does not violate the invariants in \S\ref{sec:design_overview}.
Specifically, the kernel is not allowed to switch the kernel page table by updating \code{TTBR1\_EL1}.
When the kernel attempts to update \code{TTBR0\_EL1}, if the page table is a \secure user page table, \tcb will replace it with the corresponding \shadow page table.
The \tcb also mediates updates to address space identifier (ASID) to prevent the kernel from accessing sensitive user space through cached TLB entries tagged with the user-space ASID, bypassing the need for a TLB flush.
On the other hand, if it is the first time the \tcb processes the page table, \ie, a new process, the \tcb will walk the entire page table to mark page table pages as read-only, count page references, and invalidate any mappings to protected pages, \eg, sensitive pages in clear-text, secure \extbl, and page table pages, before it returns control back to the kernel.
These are enabled by the \tcb's bookkeeping about the physical frame status (\S\ref{sec:guardian_design}).

For (\textit{ii}), the kernel invokes \tcb through \code{g\_set\_pt} as described in \S\ref{sec:impl_os}.
In a trap, \tcb walks the subtree associated with the target page table entry.
In the page table walk, \tcb marks page table pages as read-only and refuses any mappings to decrypted sensitive pages or writable mappings to secure \extbl and page table pages.
As a result, \tcb guarantees that all page table pages are read-only to the kernel even if the page tables are growing.

\paragraphb{Interrupt Table Control}
\design protects the secure \extbl in the same way as it protects the page tables.
The secure \extbl is a wrapper of the original \extbl, where each entry includes an instruction to invoke \tcb's ABI \code{g\_interrupt} before jumping to the original interrupt handling logic.
To be more precise, in ARMv8-A, we add a \code{smc} instruction as the first instruction for all entries, followed by the original interrupt handling logic.
As a result, by switching to using the secure \extbl while a \secure process is running (\S\ref{sec:secure_exception_handle}), we can invoke the \tcb to mediate control flow changes due to system calls and interrupts for the \secure processes.
We note that the kernel cannot update the \vbar while a \secure process is running.
Moreover, table switching does not require a TLB flush because \design uses different kernel virtual addresses for the two tables.

\paragraphb{Page Fault in Semantic Access}
When being invoked by \mbox{\code{g\_move\_umem}} and handling semantic accesses, \tcb may trigger page faults when it copies the user-space data on behalf of the kernel.
In this case, \tcb delegates the page fault to the kernel and applies the \trapped (\S\ref{sec:secure_exception_handle}).
More precisely, \tcb sets the return address, i.e., the \code{ELR\_EL1} register on ARMv8-A, to where the kernel makes the \code{g\_move\_umem} ABI call, so that the kernel will retry the semantic access after handling the page fault.

\paragraphb{Invoking \tcb}
\autoref{tab:tcb_abi} summarizes the scenarios and methods for invoking \tcb.
For calls originating from the OS or the interrupt table, a single \code{smc} instruction suffices to change the privilege level.
However, for invocations from sensitive processes, a direct transition from EL0 to EL3 is not feasible, as no instruction supports this transition.
To address this problem, the corresponding trampoline in a sensitive process triggers a trap into EL2 by accessing \code{CTR\_EL0} and then immediately invoke \tcb with a \code{smc} instruction.
We note that \mbox{\code{HCR\_EL2}} needs to be properly configured for this trapping.
This method leverages EL2 as a bridge to invoke \tcb.
This is quite similar to how the kernel's update to the virtual memory control is trapped into \tcb, described above.

\subsection{\Secure Process Execution}
\label{sec:impl_proc}

Next, we describe how \design works during the life cycle of a \secure process.

\paragraphb{Binary Adaptation}
\design supports legacy binaries, but must adapt them for secure execution.
\design assumes the developer and \tcb have their own public-private key pairs.
At compile time, the developer prepares the binary via the \adp.
The \adp generates symmetric keys to encrypt and sign loadable segments and metadata such as binary headers.
It encrypts the symmetric keys with \tcb's public key and embeds them into the binary along with their signatures so that \tcb can check the integrity of the keys and the segments.
We store all signatures in an added \code{signature} segment and apply the same adaptation to library binaries if the app is dynamically linked.
We also reserve a virtual memory area in the form of a \code{BSS} segment to store heap and stack signatures at runtime.

The \adp also adds a text segment with the three user-space trampolines (\autoref{tab:tcb_abi}) to the binary and redirects the entry point (i.e., \code{\_start}) to one of the trampolines (\code{g\_proc\_create}).
This \shim only contains a few instructions, smaller than the shim introduced in Overshadow~\cite{chen2008overshadow} by three orders of magnitude.
We discuss the use of each trampoline in detail in the following.

\paragraphb{\Secure Process Creation}
When the kernel transfers control to the entry point of a process as the last step of process creation, it unknowingly transfers to the \code{trampoline} segment to invoke \tcb via ABI \code{g\_proc\_create} (see \trapped in \S\ref{sec:secure_exception_handle}).
The \tcb checks the integrity of the metadata with the \adp's public key and extracts the symmetric keys with its private key.
With the headers, \tcb walks the list of virtual memory area descriptors in the kernel space to learn the virtual addresses of the segments.
Then it walks the user page table. If a page is present, \tcb hides the mapped frame from the kernel, checks its integrity, and decrypts it with the extracted keys.
Finally, \tcb configures the \vbar to the secure \exvec (\S\ref{sec:secure_exception_handle}) before returning control to the original entry point.

\paragraphb{Clone and Fork}
\tcb can identify \code{clone} and \code{fork} system calls from the stored context (\S\ref{sec:secure_exception_handle}).
The \tcb is aware that the stored context will be restored twice if the system call succeeds (for the parent and the child).
When a cloned child thread starts with \code{g\_proc\_resume}, the \tcb recognizes the thread and restores the context accordingly, based on the sensitive process identifier and the stack pointer (\S\ref{sec:guardian_design}).
As for a forked child process, we modify the kernel to allocate a new \shadow page table and invoke \tcb's ABI \code{g\_fork} by the end of the fork system call handling.
So \tcb can identify the forked process and restore context accordingly when it returns from the handling.

\paragraphb{Signal Handling}
There are two challenges in signal handling: (\textit{i}) execution environment setup and (\textit{ii}) control transfer.

For (\textit{i}), the Linux kernel reuses the user stack (or an alternate signal stack) to execute the signal handler, and thus writes to the user space with \code{copy\_to\_user}.
We add a capability starting at the top of the stack whenever an interrupt happens and destruct it before resuming process execution, since signal handling can happen at any time when the control returns from an interrupt.

For (\textit{ii}), \tcb can identify system calls like \code{sigaction} and learn the addresses of legitimate signal handlers from the parameters when the system calls are trapped in \S\ref{sec:secure_exception_handle}.
It redirects the pointer of the signal handler in parameter to the trampoline that invokes \code{g\_proc\_signal} (see \trapped in \S\ref{sec:secure_exception_handle}).
Therefore, \tcb is invoked in all attempts to run a user-defined signal handler.
At which point, it transfers control to the original signal handler after verifying the legitimacy.
As a result, \design can prevent malicious control transfer to execute \secure process execution.

\section{Evaluation}
\label{sec:eval}

We quantify \design's performance for both sensitive and nonsensitive configurations compared to that on vanilla Linux with both micro and macro benchmarks.
We focus on the run-time overhead because the overhead of secure boot and binary adaptation is one-time and small.

\begin{table}[t]
    \caption{List of micro and macro benchmarks}
    \label{tab:benchmark}
    \centering
    \begin{tabular}{ l | p{6.5cm} } 
        \hline
        \textbf{Name} & \textbf{Description} \\
        \hline\hline
        LMbench & \code{LMbench} v3.0-a9~\cite{lmbench} micro benchmarks \\
        \hline
        OTP & One-time password generator~(\code{OTP}) from open source code of Ginseng~\cite{yun2019ndss} \\
        \hline
        DNN & Deep neural network-based object classification~(\code{DNN}) from open source code of OpenCV~\cite{opencv-DNN-example} \\
        \hline
        Redis & \code{redis} v7.2.5 using the \code{memtier} benchmark v1.2.0 with default \code{redis} protocol \\
        \hline
        Memcached & \code{memcached} v1.6.9 using the \code{memtier} benchmark v1.2.0 with \code{memcache\_text} protocol \\
        \hline
        Nginx & \code{nginx} v1.20.1 server handling 100 concurrent requests from remote \code{ApacheBench} v2.3 client \\
        \hline
        Apache & \code{apache} v2.4.46 server handling 100 concurrent requests from remote \code{ApacheBench} v2.3 client \\
        \hline
    \end{tabular}
\end{table}

We run the benchmarks listed in \autoref{tab:benchmark}.
For all evaluations, we run \design on a Raspberry Pi 4 Model B, which is equipped with Quad core Cortex-A72 at 1.8GHz and 8~GB DRAM.
For those benchmarks that involve both servers and remote clients, the servers are running on a Raspberry Pi 4 Model B, while the clients are running on an Intel NUC13ANHi7. The two machines are connected directly with an Ethernet cable.

\subsection{Micro benchmark}
\label{sec:eval_micro}

\begin{figure}[t]
    \centering
    \includegraphics[width=0.4\textwidth]{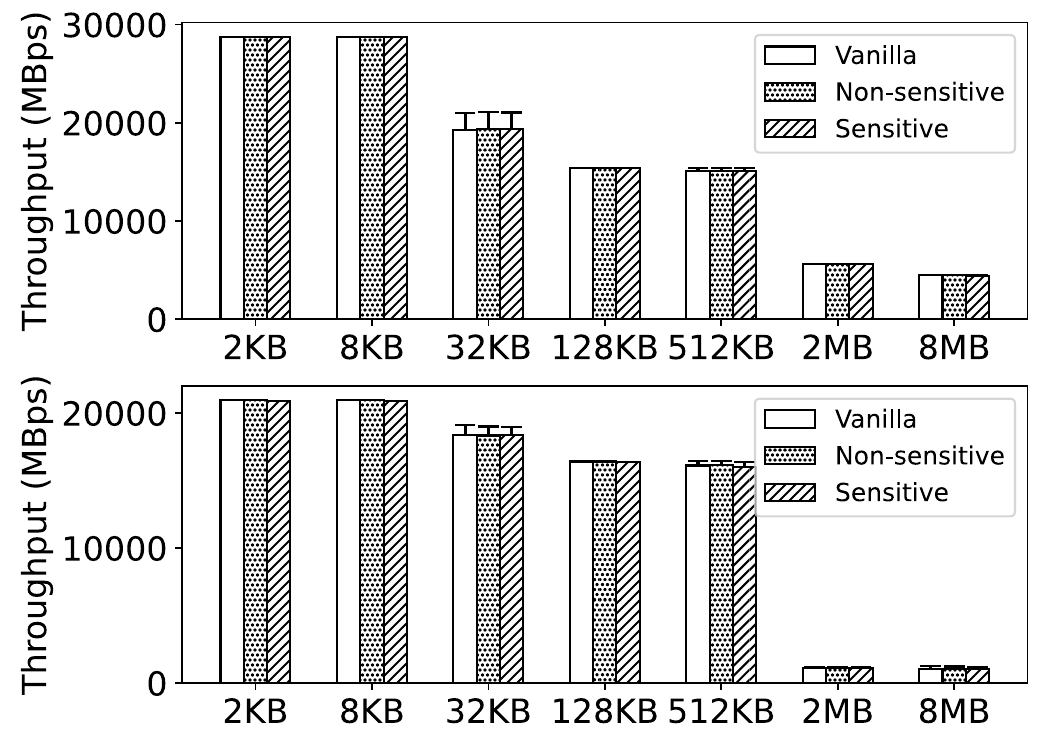}
    \caption{LMbench memory throughput benchmark: read (top) and write (bottom).
    The overhead of accessing application's own memory is negligible in \design.}
    \label{fig:mem_access}
\end{figure}

\paragraphb{Negligible overhead for application's own memory access}
\design aims to impose as little overhead as possible for the common case in which a \secure process accesses its own memory.
In \autoref{fig:mem_access}, the memory throughput benchmark from LMbench~\cite{lmbench} shows such overhead is negligible.

\begin{figure}[t]
    \centering
    \includegraphics[width=0.4\textwidth]{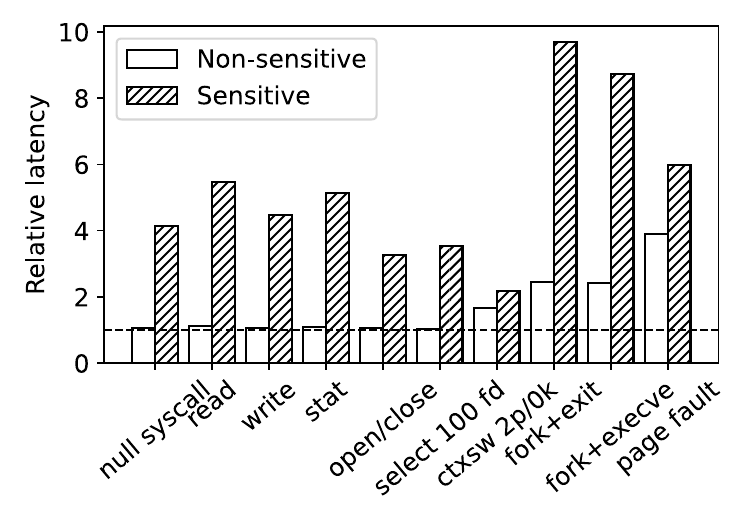}
    \caption{LMbench latency micro benchmarks.
    The two bars for each benchmark represent the latency ratio on Non-sensitive and Sensitive configurations compared to the latency on Vanilla Linux, indicated by the horizontal dashed line.}
    \label{fig:lmbench}
\end{figure}

We further analyze the sources of overhead introduced by \design using more LMbench benchmarks.
\autoref{fig:lmbench} presents the results by comparing the latency of non-sensitive and sensitive configurations with that on vanilla Linux.

\paragraphb{Overhead for system calls}
System call related benchmarks, including \textit{null syscall}, \textit{read}, \textit{write}, \textit{stat}, \textit{open/close} and \textit{select}, introduce negligible overhead to nonsensitve applications, \ie, the ratios are close to 1.
On the other hand, the latency of sensitive applications is $3$ to $5\times$, which includes the overhead of trapping the system call to store \secure context and switching tables, as well as the overhead of semantic access in system call handling.

\paragraphb{Page table update overhead anaysis}
The other four benchmarks indicate that \textit{context switch}, \textit{page fault}, \textit{fork} and \textit{execve} impose non-negligible overhead on nonsensitive applications, because all these involve trapping into \tcb for security checks.
A \textit{context switch} necessitates an update to the page table base register, which is trapped by hardware into \tcb.
For each \textit{page fault}, the kernel also must call the \tcb to update the page table.
As for \textit{fork} and \textit{execve}, \design does not trap these two events explicitly for non-sensitive processes. However, both benchmarks create a new process, which implies page table updates when creating the page table for the new process.

On the other hand, \design imposes a higher overhead on sensitive applications.
We classify the benchmarks into two groups and analyze them one by one.
First, \textit{context switch} introduces overhead due to the trapping of page table base register updates as mentioned above.
It is slightly higher than that on nonsensitive applications because \tcb has to look up its bookkeeping for the address of the cloak page table for the update.
Second, for those benchmarks involving page table updates, including \textit{fork}, \textit{execve} and \textit{page fault}, they introduce overhead due to the trappings of the page table updates, as mentioned above.
However, the overhead is much higher because (un)mappings for a \secure process involve cryptography such as encryption/decryption and signature generation/verification.
As the cryptography operations are expensive, the overhead is much higher than that on a nonsensitive process.

To conclude, the major source of overhead for non-sensitive applications is page table updates.
For sensitive applications, the overhead of page table updates and the related cryptography operations is high.
Beyond that, overhead introduced by system calls and semantic accesses is non-negligible.

\subsection{Macro benchmark}
\label{sec:eval_macro}

Micro benchmarks (\S\ref{sec:eval_micro}) indicate that \design imposes high overhead on an application when it calls \textit{fork} and \textit{execve}.
So in the following macro benchmarks, we first present and analyze the results from short-lived applications that \textit{fork} and \textit{execve} from binaries. Then we present results of the long-lived daemon applications.

\paragraphc{1) Short-lived applications}
We use the two open source applications, \ie, one-time password generator (\code{OTP} from Ginseng~\cite{yun2019ndss}) and DNN-based object classifier (\code{DNN} from OpenCV~\cite{opencv-DNN-example}), for our evaluation. These are popular applications in real world scenarios and both are sensitive as the users care about the confidentiality of the input/output.

\paragraphb{Cryptography is the major source of overhead for short-lived application}
The \code{Original} columns in \autoref{tab:latency_otp_dnn} presents the results of \code{OTP} and \code{DNN} benchmarks, measuring the average latency of 1000 executions with \code{time.perf\_counter} from Python.
The overhead of non-sensitive \code{OTP} and \code{DNN} are $65.6\%$ and $22.8\%$ respectively, while the latency of sensitive ones are $2.8\times$ and $9.1\times$.
The reason why the sensitive \code{OTP} has less overhead, \ie, $2.8\times$ instead of $9\sim10\times$, is because \code{OTP} is compute-intensive and does not require much memory.
So it does not trigger as many page faults and cryptography operations as \code{DNN}.
In contrast, a cold-started \code{DNN} consumes more pages and necessitate page table updates that involve expensive cryptography.

\paragraphb{Performance optimization for short-lived applications}
The high overhead of cryptography imposes a formidable cost to create a \secure process: the cold start of the program causes a lot of page faults (and cryptographic operations).
This suggests that \design is more efficient at protecting long-lived services than ephemeral ones.
For example, our original implementations (\code{Original}) of \code{OTP} and \code{DNN} create a new process for each authentication and inference request, respectively. 
After we slightly modify them into long-lived applications (\code{Optimized}), the overhead from \design drops, as shown in \code{Optimized} columns in \autoref{tab:latency_otp_dnn}.

\begin{table}[t]
    \caption{Latency of \textbf{OTP} and \textbf{DNN}. The three rows represent latency of vanilla, non-sensitive, and sensitive executions. The ``Original'' and ``Optimized'' columns represents latency of executions before and after optimizing the short-lived applications into daemons.}
    \label{tab:latency_otp_dnn}
    \centering
    \begin{tabular}{ | l | l | l | l | l | } 
        \hline
        \multirow{2}{*}{Benchmarks} & \multicolumn{2}{|c|}{\textbf{OTP} ($\mu$s)} & \multicolumn{2}{|c|}{\textbf{DNN} (ms)} \\
        \cline{2-5}
        & Original & Optimized & Original & Optimized \\
        \hline\hline
        Vanilla & 2681.58 & 43.27 & 1113.82 & 581.64 \\
        \hline
        Non-sensitive & 4442.00 & 46.43 & 1368.04 & 597.51 \\
        \hline
        Sensitive & 7399.24 & 62.51 & 10121.84 & 640.60 \\
        \hline
    \end{tabular}
\end{table}

\begin{figure}[t]
    \centering
    \includegraphics[width=0.38\textwidth]{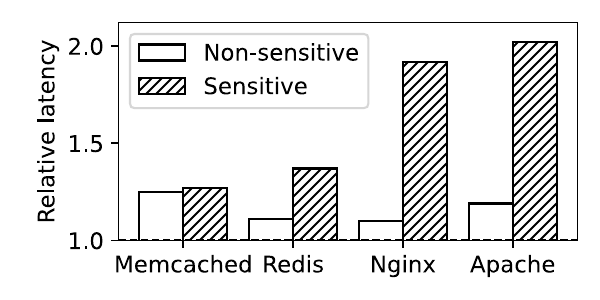}
    \caption{Long-lived application macro benchmarks.
    The two bars for each benchmark represent the latency ratio on Non-sensitive and Sensitive configurations compared to the latency on Vanilla Linux, indicated by the x-axis (y-tick starts from 1).}
    \label{fig:daemon}
\end{figure}

\paragraphc{2) Long-lived applications}
We use four applications that are popular in real-world scenarios for our long-lived application benchmarks.
\code{Memcached} and \code{Redis} are in-memory key-value databases that are memory intensive, while \code{Nginx} and \code{Apache} are web servers that are I/O intensive.

\autoref{fig:daemon} presents the relative latency of nonsensitive and sensitive configurations, noting that the x-axis represents the latency in vanilla Linux and the y-tick starts from 1.
We also note that even if we use the same version of \code{memtier} to benchmark both \code{Memcached} and \code{Redis}, the parameters are different since the two applications use different protocols.
The results show the overhead varies from one application to another and we next analyze them case by case.

\paragraphb{Long-lived applications trigger few cryptography operations}
The overhead of both sensitive and non-sensitive \code{memcached} are high but close, which implies that the major source of overhead is page table updates that do not involve cryptography.
Unlike a cold-started application, most pages of a long-lived application are already decrypted.
As a result, most pages will remain decrypted as long as the kernel does not perform non-semantic access.
To be more precise, while there are not many non-semantic accesses such as page swapping,
(\textit{i}) the kernel allocates anonymous pages to the application as its working set grows large, or the kernel reclaims page when the application frees the pages;
(\textit{ii}) the kernel may migrate pages via secure ABI.
Neither of the cases involves encryption/decryption.

\paragraphb{System calls are the major source of overhead for long-lived application}
While both \code{Memcached} and \code{Redis} are in-memory key-value databases, they do not share the same overhead pattern. We investigated the execution of \code{Redis} and found that it makes frequent system calls. Surprisingly, about half of the system calls are \code{gettimeofday}. We note that we are not the first to discover this issue, as it is already reported in Redis’s GitHub repository~\cite{redis_gettimeofday, redis_freshtime}.
Such system calls are not necessary for in-memory key-value databases since they do not require a very fresh time~\cite{redis_freshtime}. This suggests that the higher overhead of \code{Redis} can be optimized away.

On the other hand, as IO-intensive web servers, \code{Nginx} and \code{Apache} make more frequent system calls, such as \code{epoll\_wait} and \code{recv\_from}, which is the main reason why they both have a much higher overhead in the sensitive configuration.

\begin{figure}[t]
    \centering
    \includegraphics[width=0.46\textwidth]{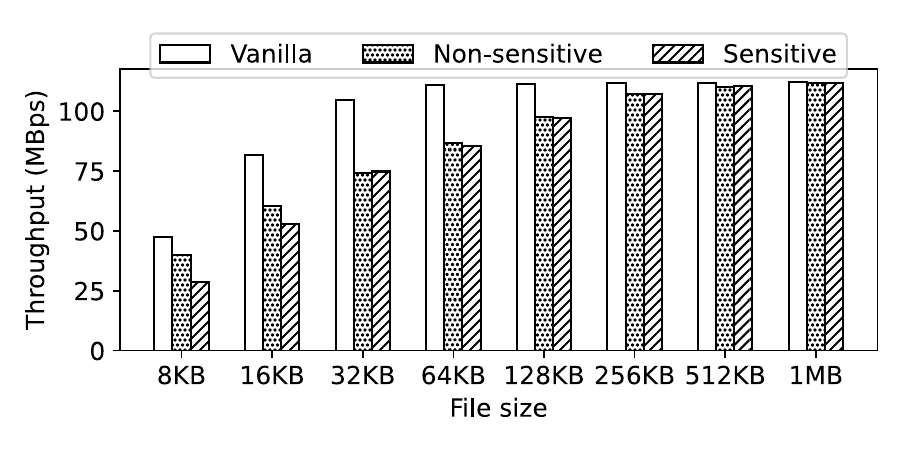}
    \caption{\program{Nginx} throughput for serving an HTML file repeatedly over a local-area network (Ethernet).
    Since most of the overhead is fixed, the relative overhead of \design diminishes as the file size increases.}
    \label{fig:eval_nginx}
\end{figure}

\paragraphb{Performance optimization for long-lived applications}
Since system calls are a major source of overhead for long-lived applications, a straightforward optimization is to eliminate unnecessary system calls, such as the frequent use of \code{gettimeofday} in Redis~\cite{redis_gettimeofday, redis_freshtime}.

Beyond that, another optimization is to batch multiple system calls into one, if possible, or to fully utilize each system call.
For example, sending a large file instead of several smaller ones with \code{Nginx} is more efficient.
\autoref{fig:eval_nginx} shows that the overhead diminishes as the file size increases.
We note that in the evaluation shown in \autoref{fig:eval_nginx}, we do not encrypt the file to better demonstrate the overhead imposed by \design.

\begin{table*}[t!]
\begin{minipage}{.48\textwidth}
\caption{Raspberry Pi 4 Model B: Overhead in \# of cycles for CPU privilege mode change from Row to Column.}
\label{tab:mode_switch_rpi}
\centering
\small
 \begin{tabular}{c||c c c c} 
 Mode & EL0 & EL1 & EL2 & EL3 \\
 \hline\hline
 EL0 & & 262  & 975  & N/A \\ 
 EL1 & 59  & & 1046 & 273  \\
 EL2 & 65 & 59  &  & 307 \\
 EL3 & 66  & 58  & 283  & \\
 \hline
 \end{tabular}
\end{minipage}
\hfill
\begin{minipage}{0.48\textwidth}
\caption{Hikey 960: Overhead in \# of cycles for CPU privilege mode change from Row to Column.}
\label{tab:mode_switch_hikey960}
\centering
\small
 \begin{tabular}{c||c c c c} 
 Mode & EL0 & EL1 & EL2 & EL3 \\
 \hline\hline
 EL0 & & 44  & 868  & N/A \\ 
 EL1 & 31  & & 1308 & 43  \\
 EL2 & 248 & 16  &  & 817 \\
 EL3 & 17  & 36  & 634  & \\
 \hline
 \end{tabular}
\end{minipage}
\end{table*}

\paragraphb{Performance Comparison with Related Systems}
The most related recent systems are BlackBox~\cite{van2022blackbox} and TrustShadow~\cite{guan2017trustshadow}.
Unfortunately, an apple-to-apple performance comparison with any of them turns out to be rather difficult.
First, they are closed-source and/or outdated, thus a fair direct comparison is not practically possible. 
Second, the systems require different hardware due to their varied goals and implementations, preventing us from comparing them under the same controlled hardware and environment setup.
We will discuss the differences in the next section (\S\ref{sec:related}).

\subsection{Hardware impact on performance}
\design's overhead highly depends on the silicon, because it involves frequent switching among three privilege levels for user, kernel, and \tcb, respectively.
We devise a nano benchmark to measure the latency of such privilege level switching on Raspberry Pi 4 Model B, the system used in the reported evaluation, and Hikey 960, a system used in an early prototype of \design reported in~\cite{li2022mprotect}.
We find that the cost of switching can be very different, as shown in \autoref{tab:mode_switch_rpi} and \autoref{tab:mode_switch_hikey960}.
In particular, switching involving EL2 and EL3 is very expensive, which to some extent explains the high overhead introduced by system calls in our implementation (\autoref{fig:lmbench}).
Moreover, because there is no instruction that can directly trigger a trap from EL0 to EL3 on ARMv8-A, our implementation uses EL2 as a stepping stone into \tcb (in EL3) from a \secure process (\S\ref{sec:impl_tcb}).
This further contributes to the overhead.
All these suggest that we should implement \tcb in EL1 using software-based privileges such as those featured in~\cite{azab2016skee,dautenhahn2015nested}.

\subsection{Kernel functionality}
\label{sec:eval_coverage}

In this section, we show that the Linux kernel can fulfill its functions under the restrictions imposed by \design.
Not all of these functions are supported in related systems.

\paragraphb{System Call Coverage}
We evaluate system call coverage in \design with the Linux Test Project (version 20230516)~\cite{linux_test_project} test cases for system calls.
We run the adapted benchmark binaries as sensitive applications in \design and compare the results with that of running the vanilla benchmarks on vanilla Linux.
There are 1340 test cases, 197 of which are not supported by vanilla Linux on Raspberry Pi 4 due to different kernel configuration or missing hardware/OS features.
On vanilla Linux, 1065 test cases pass, while the remaining 78 fail.
With \design, 1035 test cases pass, which is more than 95\% of those that also pass on vanilla Linux. In comparison, BlackBox passes about 90\% of those that pass on vanilla Linux~\cite{van2022blackbox}, while TrustShadow did not report their coverage.

We analyze the system calls that \design failed to support and identify the reasons for these failures. In summary, these system calls either require trust beyond the boundary of a single process (\eg, identifiers for shared memory regions) or rely on kernel/driver-specific knowledge.

\begin{itemize}[topsep=0.1em, leftmargin=*]
    \item \code{process\_vm\_readv} and \code{process\_vm\_writev}:~ These two system calls transfer data between processes identified with a provided \code{pid}, while \design cannot trust the \code{pid} as an identifier (\S\ref{sec:guardian_design}).
    \item \code{shmat} and its relatives:~ These system calls use \code{shmid} to identify a piece of shared memory, which is neither trusted nor recognized by \design.
    \item \code{ioctl}:~ Blindfold cannot fully support it as it relies on the device driver-specific semantics.
    \item \code{io\_uring} and \code{vmsplice}:~ We defer supporting system calls that require either very complicated internal structures such as pointer in pointer, especially more than 3-levels, or in-kernel state that is not known to the Guardian.
    \item We omitted other minor test cases as they do not significantly affect Linux’s functionality, \eg, \code{profil} system call.
\end{itemize}

\paragraphb{Non-semantic Access}
We evaluate non-semantic access using the \code{migrate\_pages} system call.
\code{migrate\_pages} triggers the OS to migrate all pages of a process from one memory region to another.
Because the \code{migrate\_pages} handler avoids moving pages within the same memory region, we modify the kernel to bypass this optimization since the Raspberry Pi 4 does not have multiple memory regions (NUMA).

In more than 1,000 page migrations, all page movements are successful, and the sensitive process continues to run correctly.
We note that recent related systems, such as TrustShadow~\cite{guan2017trustshadow} and BlackBox~\cite{van2022blackbox}, do not support the non-semantic access necessary for page migration because they hide memory allocated to secure processes/containers from the OS.

\paragraphb{Supporting Existing Binaries with Adaptation}
\design supports existing binaries only requiring adaptation (\S\ref{sec:impl_proc}).
We adapted all the applications for micro and macro benchmarks used in our evaluation.

\section{Related Work}
\label{sec:related}

\paragraphb{Virtualization-based Solutions}
\textit{Overshadow}~\cite{chen2008overshadow} employs a hypervisor as the TCB to control user data access from the untrusted OS using nested virtualization (Figure \ref{fig:design_comparison}).
Although often justly criticized for its massive TCB~\cite{van2022blackbox}, Overshadow pioneered two ideas.
First, goal-wise, it provides the OS an encrypted view for non-semantic access to user space.
Although many later systems eschew this goal~\cite{guan2017trustshadow,van2022blackbox}, having an encrypted view is essential to allow resource management inside the OS for a protected process.
Second, solution-wise, it pioneered the use of nested virtualization to protect applications from the underlying untrusted OS, which is adopted by many later systems~\cite{li2019hypsec,li2021twinvisor,van2022blackbox}.
While \design shares the same concept of encrypted view, \design uses switching, instead of nesting (\S\ref{sec:insight}).
Most recently, \textit{BlackBox}~\cite{van2022blackbox} further innovates the use of nested page tables by substantially reducing the TCB running in the hypervisor mode and expanding the protection to containers.
BlackBox, however, does not allow non-semantic accesses by the OS; once a physical memory frame is allocated to a protected container, it disappears from the OS's view.
As a result, many OS functions will stop working for these frames, \eg, swapping, memory compression~\cite{linux_mem_compression}, and page migration including related signals such as \code{move\_pages} and \code{migrate\_pages}.

\paragraphb{Hardware-based Solutions}
Instead of using the hypervisor mode for TCB, another line of work takes advantage of architecture-specific hardware features.
\textit{TrustShadow}~\cite{guan2017trustshadow} runs a protected process inside the ARM TrustZone's Secure World that is isolated from the OS in the Normal World by hardware.
The memory used by the protected application is allocated in the Secure World and as a result, disappears from the OS, just like in BlackBox.
Other hardware-based solutions, such as Intel SGX, depend on architecture-specific hardware implementations, making them difficult to deploy on other architectures~\cite{mckeen2013innovative,xing2016intel}.

Without systematically addressing access to user space by the OS, these works rely on case-by-case solutions that are incomplete and inefficient.
For instance, since the OS cannot access the user space, the TCB needs to prepare a buffer to store syscall arguments so that the OS access them from the buffer. As a result, the TCB not only copies data between the OS and the user application, but also manages the memory used by the buffer, which can be arbitrary large for some system calls, \eg, \code{read()/write()}. In addition, buffer-based data copy cannot support \code{futex} which requires atomic operation of data copy and wait queue update. Existing work proposed another workaround to address this issue, \eg, a \code{futex} syscall handler inside the TCB~\cite{chen2008overshadow} and modification to the \code{futex} handler in the OS~\cite{van2022blackbox}.

\paragraphb{Paravirtualization-based Solutions}
\design bears similarity to \textit{paravirtualization} in which a VMM works with a modified guest OS and/or augmented applications making hypercalls. However, unlike the existing VMMs that protect user-space, \eg, \textit{InkTag}~\cite{hofmann2013inktag} and \textit{Sego}~\cite{kwon2016sego}, \design pursue different goals resulting in different design choices: (1) reducing the TCB size, much smaller than existing VMMs, and (2) preserving optimizations in modern OSes such as page migration, instead of nesting page tables via virtualization, which effectively nullifies the OS-level optimizations.

\begin{figure}[!t]
    \centering
    \includegraphics[width=0.48\textwidth]{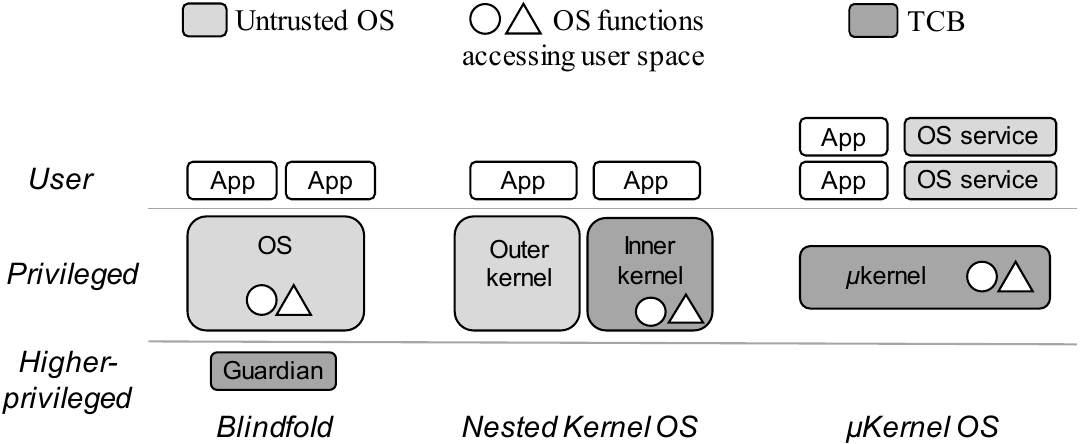}
    \caption{Nested kernel and microkernel-based OSes limit OS access to the user space by moving OS functions outside the TCB.
    However, OS functions requiring user-space access must reside in the TCB, increasing the TCB size.}
    \label{fig:related_work}
\end{figure}

\paragraphb{Comparison with Related OS Designs}
\textit{Exokernel}~\cite{engler1995exokernel} and \textit{$\mu$Kernel} OSes can also prevent OS services from accessing user memory by relocating them to the user space. However, without answering how OSes can manage memory and serve system calls without unfettered access to the user space, these functionalities have to stay in the TCB (\autoref{fig:related_work}).
\textit{Nested Kernel}~\cite{dautenhahn2015nested} provides a mechanism to implement privilege in software. \design can use such a mechanism in place of the architecture requirement (A1 in \S\ref{sec:assumption}).
Indeed, even though Nested Kernel does not aim at protecting user address space, one can disable the outer kernel from accessing the user space and let the nested kernel control page-table updates to protect the user space. However, without access to the user space, the outer kernel cannot anymore manage user pages or handle system calls, resulting in a larger inner kernel or TCB as shown in \autoref{fig:related_work}.

\paragraphb{Discussion on TEE and Confidential VMs}
Compared to TEEs that rely on hardware features such as ARM TrustZone, \design offers a software-driven solution with broader applicability. In particular, certain concepts from \design could also enhance existing hardware-based solutions. For example, a Realm management monitor in ARM CCA~\cite{arm-cca} can apply \design’s capability system design to serve the VMs in Realm. In addition, since \design is orthogonal to confidential VM solutions such as Intel TME-MK~\cite{intel-tdx}, \design can improve security \textit{inside} a confidential VM, between the OS and applications.

\section{Discussion}
\label{sec:discussion}

We next discuss (1) how we implement a basic x86-64 port (without full features), (2) some unsolved problems in \design design and new opportunities.

\paragraphb{Support for x86-64}
We have implemented \design on x86-64 as a proof of concept and report the two key differences from that on ARMv8-A.
The x86-64 \tcb is implemented as an extension of the micro hypervisor Bareflank~\cite{bareflank}.
Although our x86-64 prototype uses the Host mode to run \tcb, we do not use nested paging or EPT.
Next, we present some key differences from the ARMv8-A prototype.

First, x86 requires a different hardware trapping mechanism, as mentioned in \S\ref{sec:background} and summaried in \autoref{tab:arch_req}.
To trap virtual memory control on x86-64, \tcb sets the \textit{CR3-load exiting} bit and the \textit{MSR bitmaps} of \code{EFER} in Virtual Memory Control Structure (VMCS), as well as configures the \textit{Guest/Host Masks} and \textit{Read Shadows} for \code{CR0} and \code{CR4} accordingly.

Second, x86 requires a different interrupt mechanism because it has a more strictly structured interrupt table.
On ARMv8-A, the interrupt/exception table consists of 16 entries and each entry contains up to 32 instructions.
On x86-64, an interrupt table is a jump table where each entry is a single 64 bit address indicating the target interrupt handler.
Therefore, unlike ARMv8-A, we cannot directly insert a guardian call into each entry.
We resolve this issue with indirection.
Specifically, the x86-64's \emph{secure \extbl} points to a ``wrapper table'' of which each entry contains only two operations, a guardian call, \ie, calling the \tcb's ABI \code{g\_interrupt}, and a jump instruction back to the original interrupt handler.

One complication arises as x86-64 supports fast system calls with the \code{syscall} instruction, which jumps to the system call handler pointed by the \code{lstar} register, bypassing the \extbl.
We solve it by modifying the \code{lstar} register to point to another wrapper, which jumps to the original address stored in \code{lstar} after invoking the \tcb.

Although not all features, such as secure ABI for optimizing non-semantic accesses, have yet been supported in our x86-64 prototype, we believe it adequately demonstrates that \mbox{\design’s} design can be applied to x86-64.

\paragraphb{Formal verification of the implementation} Although the \name design provides security properties as shown in \S\ref{sec:attack_surface}, the implementation may contain deviations from the design that can lead to a new attack surface. Formal verification methods can be used to address this problem by mechanically checking for gaps between the design specification and the implementation~\cite{klein09sel4, gu16certikos}. Since formal verification involves significant overheads in many cases, especially in terms of developer hours required to describe specifications and invariants, we leave this for future work.

\paragraphb{Side Channel}
In this paper, we have not considered side-channel attacks in \design. There are two side channels that we should investigate. 
First, the OS still knows what system calls are made and with what arguments. It may be able to infer about a \secure process based on the system calls made.
Second, while the OS cannot modify a page table or read a user page in \design, it can infer about which user pages are accessed and when because it still handles page faults.

\paragraphb{Memory Compression}
The OS requires semantic accesses to memory in order to compress them effectively. The encrypted view supported by \design will render compression ineffective. As a result, memory compression~\cite{linux_mem_compression} is the only kernel memory optimization that loses its effectiveness under \design, although it still works. To support memory compression in \design, the application can explicitly grant the OS the capability to gain semantic access to the pages to be compressed. This requires the application developer to modify their applications accordingly. 

\paragraphb{Optimizing \design for I/O-intensive Workload}
We shows with \program{Nginx} (\autoref{fig:eval_nginx}) that \design introduces a fixed overhead per system call independent of the size of data being served. Therefore, I/O-intensive applications may want to amortize the overhead, for example, by using multi-message system calls such as \code{recvmmsg} or \code{sendmmsg}. However, \design currently requires the OS to use one \tcb ABI call to copy each packet into the protected memory of the process. One way to overcome this limitation is to improve the \tcb ABI to support batching. With that, the OS can employ the scatter-gather pattern to collect multiple packets per \tcb ABI call.

\section{Conclusion}

Modern operating systems (OSes) assume that applications trust them, granting the OS unfettered access to any data in user applications. \design demonstrates that such unrestricted memory access is not fundamentally necessary for the OS to perform its tasks, including memory management.
We implemented a prototype of \design on ARMv8/Linux, leveraging a tiny trusted program, called \tcb, running at a higher privilege level to mediate memory access and exception handling by the OS. 
We evaluated \name’s performance using macro and micro benchmarks, observing that \name provides competitive performance with a runtime TCB that is 2 to 3$\times$ smaller compared to previous work.

\section*{Acknowledgments}
\noindent This work is supported in part by NSF Awards \#1730574, \#2130257, and \#2112562. The authors are grateful to Dr. Min Hong Yun for his contribution during the early stage of the project.



%
\newpage
\bibliographystyle{IEEEtran}
\bibliography{abr-long,main}

\newpage
\appendix
\section{Appendix}
\label{sec:casestudy}


We conduct a case study of why and how the OS accesses user space using the Linux kernel.
We note that others have studied user-space access by Linux kernel in the context of protecting the kernel from TOCTTOU attacks, e.g., most recently Midas~\cite{bhattacharyya2022midas}.
The focus of that literature is on kernel vulnerabilities due to malicious users, not protecting users from an untrusted kernel.

\subsection{Non-semantic access}
\label{sec:nonsemantic}
The vast majority of kernel's accesses to user space are concerned with the kernel moving user-space data.
In these cases, the kernel does not need the semantics of the data being moved. 
We call such access non-semantic.

\paragraphb{\code{read/write} syscalls}
This pair of syscalls are extensively used for data exchange with I/O, including the filesystem. The kernel is simply responsible for copying data between memory region pointed by the \code{buf} argument and the file.

\paragraphb{Demand paging}
In the case of a page fault because the kernel loads file-backed pages on demand, the kernel calls the registered \code{fault} file operation to let the device driver prepare the frame, then updates the user page table to map the prepared physical frame to the user address space.

\paragraphb{Swapping}
When the number of free physical frames is low, the kernel may swap out some cold user pages from the physical memory to the external storage like disk and invalidate the corresponding user page table entries.
When the user process accesses a swapped out page, it triggers a page fault.
The kernel swaps in the page to the physical memory, remaps it to the user address space, and lets the process continue.

\paragraphb{Memory compression~\cite{linux_mem_compression}}
Instead of swapping user pages to the external storage, the kernel can compress them after invalidating the associated page table entries.
When next time the user process access a compressed page, the kernel decompresses the page and remaps it to the user address space.

\paragraphb{Page migration}
To reduce the latency of accessing memory in a NUMA system or to mitigate the problem of memory fragmentation, the kernel may migrate the content of a physical frame to another and update the associated user page tables accordingly.
Linux provides two syscalls, \code{move\_pages} and \code{migrate\_pages}, to move user pages among nodes.

\subsection{Semantic access}
\label{sec:semantic}

In some cases, the Linux kernel does need to understand the user-space data it accesses. We say such accesses are \emph{semantic}.
All semantic accesses share three properties.
First, they are \emph{well-defined} in spatial (where) and temporal (when) boundaries.
Second, the user process knows when and where the kernel access its address space. 
Third, the kernel reads/writes the user space through well-defined interface, namely, \code{copy\_to\_user} and \code{copy\_from\_user}. These properties are the foundations to our solution of semantic access (See \S\ref{sec:clearview}).

\paragraphb{Syscall argument passing}
The most common cases are argument passing in syscalls. For example, the kernel must understand \code{pathname} in an \code{open} syscall to open the file. 
In these cases, the kernel accesses the user space \emph{during} the syscalls (when) and in the region defined by the arguments (where).

Most syscalls pass an argument in a single transaction. That is, the kernel copies the user data specified by the argument to the kernel space when it handles the syscall. The kernel does not keep accessing the user data afterwards.
Prior works leverage this and copy the argument data into a buffer managed by the TCB at call time for the kernel to access later.

The \code{futex} syscall, however, is a notable exception that does not work with the buffer-based approach. When the futex syscall is invoked, the kernel syscall handler reads the user-space word specified by the \code{futex} in order to determine if and how the in-kernel wait queue for the \code{futex} should be updated. The kernel considers reading the \code{futex} and updating its wait queue as a critical section that must be done atomically, because other threads may update the \code{futex} concurrently. 

\paragraphb{Robust \code{futex}}
When a thread terminates unexpectedly while holding a \code{futex}, other threads waiting for the \code{futex} may end up waiting forever. Robust \code{futex}~\cite{robust_futex} solves this problem with a collaboration between \program{glibc} and the kernel. \program{glibc} creates and manages a list of all \code{futex}es held by the thread. The thread uses the syscall \code{set\_robust\_list} to inform the kernel where the head of the list is. When the thread terminates, the kernel traverses this list starting from its head.

\paragraphb{\code{clone} syscall}
When a user process makes a \code{clone} syscall to create a new process or a new thread, it can set the \code{CLONE\_CHILD\_CLEARTID} flag (\code{flags}) and pass the address of the child thread identifier (\code{child\_tid}) to the kernel.
When the child exits, the kernel will clear the child thread identifier by writing to the address specified by \code{child\_tid}.
Another syscall \code{set\_tid\_address} is similar.

\paragraphb{Call stack write}
In two cases, the kernel must write to the user-space memory defined by the call stack. First, when creating a process from a binary, the kernel must prepare its call stack by placing arguments at the bottom of the stack.
Second, during signal handling, the kernel prepares the call stack (or signal stack) before handing control to a user-space signal handler.
The signal handler can either use the call stack or an alternate \emph{signal stack} (also in the user space), defined by the \code{sigaltstack} syscall when the signal handler is being established using the \code{sigaction} syscall.
In both cases, the semantic write is well-defined in time and space.

\end{document}